\documentclass[11pt]{article}

\usepackage{epsfig,amssymb}

\newcommand{\qnc}{\sf QNC^{\rm 0}_f}
\newcommand{\qac}{\sf QAC^{\rm 0}_f}
\newcommand{\qtc}{\sf QTC^{\rm 0}_f}
\newcommand{\qnck}{\sf QNC^{\it k}_f}
\newcommand{\qack}{\sf QAC^{\it k}_f}
\newcommand{\qtck}{\sf QTC^{\it k}_f}
\newcommand{\nc}{\sf NC^{\rm 0}}
\newcommand{\ac}{\sf AC^{\rm 0}}
\newcommand{\tc}{\sf TC^{\rm 0}}

\newtheorem{lemma}{Lemma}
\newtheorem{theorem}{Theorem}
\newtheorem{corollary}{Corollary}

\begin{document}

\title{Collapse of the Hierarchy of Constant-Depth Exact Quantum
Circuits}

\author{Yasuhiro Takahashi and Seiichiro Tani\\
NTT Communication Science Laboratories, NTT Corporation\\
{\small\tt \{takahashi.yasuhiro,tani.seiichiro\}@lab.ntt.co.jp}}

\date{}

\maketitle

\begin{abstract}
 We study the quantum complexity class $\qnc$ of quantum operations
 implementable exactly by constant-depth polynomial-size quantum
 circuits with unbounded fan-out gates (called $\qnc$ circuits). Our
 main result is that the quantum OR operation is in $\qnc$, which is an
 affirmative answer to the question of H{\o}yer and {\v S}palek. In
 sharp contrast to the strict hierarchy of the classical complexity
 classes: $\nc \subsetneq \ac \subsetneq \tc$, our result with H{\o}yer
 and {\v S}palek's one implies the collapse of the hierarchy of the
 corresponding quantum ones: $\qnc=\qac=\qtc$. Then, we
 show that there exists a constant-depth subquadratic-size quantum
 circuit for the quantum threshold~operation. This implies the size
 difference between the $\qnc$ and $\qtc$ circuits for implementing the
 same quantum operation. Lastly, we show that, if the quantum Fourier
 transform modulo a prime is in $\qnc$, there exists a polynomial-time
 exact classical algorithm for a discrete logarithm problem using a
 $\qnc$ oracle. This implies that, under a plausible assumption, there
 exists a classically hard problem that is solvable exactly by a $\qnc$
 circuit with gates for the quantum Fourier transform.
\end{abstract}

\section{Introduction and Summary of Results}

Quantum computers are expected to solve some problems much faster than
classical computers (e.g.\ Shor's factoring algorithm \cite{Shor}). It
is, however, still difficult to realize a quantum computer that can
perform quantum algorithms for a reasonably large input size. A major
obstacle to realizing a quantum computer is that, even if we can prepare
many qubits, we can use them only for a short time due to the coherence
time. In order to use such fragile qubits effectively, it is important
to understand the possibilities and limitations of using them. This
motivates us to study the computational power of quantum circuits with a
small amount of computation time
\cite{Moore,Green,Fenner,Hoyer,Fang,Bera,Bera2}.

In this paper, we focus on the theoretical analysis of the computational
power of constant-depth polynomial-size quantum circuits, which allows
us to analyze that of polylogarithmic-depth ones. The elementary gates
are one-qubit, CNOT, and unbounded fan-out gates. The unbounded fan-out
gate is an analog of the classical one normally assumed to be an
elementary gate for the theoretical study of classical circuits 
\cite{Vollmer}. The gate on $n+1$ qubits makes $n$ copies of a classical
source bit in a superposition and, in particular, the gate on two qubits
is a CNOT gate. It is theoretically interesting to deal with the gate as
an elementary gate since the use of the gate clarifies many differences
between quantum and classical circuits \cite{Green,Hoyer} and connects
the quantum circuit model with the one-way model~\cite{Browne}.

There are three important settings for studying constant-depth classical
circuits. All the settings allow the use of (classical) unbounded
fan-out gates. The first setting deals with constant-depth
polynomial-size classical circuits consisting of NOT gates and OR and
AND gates with {\em bounded} fan-in. The classical complexity class
$\nc$ is the class of problems solvable by (uniform families of) the
classical circuits in the setting. The second setting is the first one
augmented with OR and AND gates with {\em unbounded} fan-in, which
defines the class $\ac$. The third setting is the second one augmented
with threshold gates with {\em unbounded} fan-in, which defines the
class $\tc$. The threshold gate implements the threshold function that
outputs the bit representing whether the Hamming weight of the input is
less than a pre-determined threshold. These classes form a strict
hierarchy: $\nc \subsetneq \ac \subsetneq \tc$~\cite{Furst,Vollmer}.

Some authors consider the quantum counterparts of the above settings
\cite{Moore,Green,Hoyer}. Although it is difficult to determine what the
correct counterparts are, we regard the following settings as the
counterparts \cite{Green}, where all the settings allow
the use of unbounded fan-out gates. The first setting deals with
constant-depth polynomial-size quantum circuits consisting of one-qubit
and CNOT gates. The quantum complexity class $\qnc$, which corresponds
to $\nc$, is the class of quantum operations implementable exactly by
(uniform families of) the quantum circuits in the setting (called $\qnc$
circuits). The second setting is the first one augmented with a quantum
version of OR gates with {\em unbounded} fan-in, which defines the class
$\qac$, corresponding to $\ac$. The third setting is the second one
augmented with a quantum version of threshold gates with {\em unbounded}
fan-in, which defines the class $\qtc$, corresponding to $\tc$. It holds
that $\qnc\subseteq \qac = \qtc$ \cite{Hoyer}.

First, in order to study the relationship between $\qnc$ and $\qac$, we
consider the question posed by H{\o}yer and {\v S}palek \cite{Hoyer} as
to whether an $O(1)$-depth poly$(n)$-size quantum circuit can be
constructed for the quantum operation OR$_n$, which computes the OR
function on $n$ bits. They showed that there exists an
$O(\log^*n)$-depth $O(n\log n)$-size quantum circuit. It is a repetition
of the OR reduction, which is represented as an $O(1)$-depth circuit
that exactly reduces the computation of the OR function on $n$ bits to
that on $O(\log n)$ bits. Based on their work, we give an affirmative
answer to the question:

\begin{theorem}
There exists an $O(1)$-depth $O(n\log n)$-size quantum circuit for {\rm
 OR}$_n$.
\end{theorem}
Theorem~1 immediately implies that OR$_n$ is in $\qnc$ and thus
$\qnc=\qac$. Since $\qac=\qtc$ as described above, the hierarchy of
$\qnc$, $\qac$, and $\qtc$ collapses, i.e., $\qnc=\qac=\qtc$. This is a
sharp contrast to the strict hierarchy of the corresponding classical
classes: $\nc \subsetneq \ac \subsetneq \tc$. More generally, Theorem~1
with H{\o}yer and {\v S}palek's result immediately implies that the
hierarchy of polylogarithmic-depth exact quantum circuits collapses,
i.e., $\qnck=\qack=\qtck$ for any integer $k \geq 0$, where $\qnck$,
$\qack$, and $\qtck$ are defined similarly to $\qnc$, $\qac$, and
$\qtc$, respectively, except that they deal with $O(\log^k n)$-depth
circuits in place of $O(1)$-depth ones.

Our idea for constructing the circuit is that, after we apply H{\o}yer
and {\v S}palek's OR reduction, we compute the OR function on $O(\log
n)$ bits in depth $O(1)$ and with size exponential in $\log n$. The
exponential-size circuit is based on the representation of the OR
function as an ${\mathbb R}$-linear combination of exponentially many
parity functions. The proof of Theorem~1 depends on the fact that, in 
the $\qnc$ circuit, an unbounded fan-out gate can be used as a parity
gate \cite{Green}, which implements the parity function. We note,
however, that the relationship $\qnc=\qac$ cannot be derived only
from the computational power of parity gates in the corresponding
classical circuit, i.e., in the $\nc$ circuit. This is because, even if
the parity gates are allowed in the $\nc$ circuit, the OR function is
not in $\nc$~\cite{Hoyer}.

Second, we apply Theorem~1 to studying the relationship between $\qnc$
and $\qtc$ in detail. To do this, we consider the problem of
constructing an $O(1)$-depth small-size quantum circuit for the quantum
threshold operation TH$_n^t$, which computes the threshold function with
a threshold $t$ on $n$ bits. Theorem~1 simply yields an $O(1)$-depth
$O(tn\log n)$-size quantum circuit for TH$_n^t$ with $1 \leq t \leq
\lceil n/2 \rceil$ and an $O(1)$-depth $O((n-t+1)n\log n)$-size circuit
with $\lceil n/2 \rceil \leq t \leq n$. We show that, using Theorem~1,
for any $t$ such that the minimum of $t$ and $n-t$ is non-constant,
there exists a smaller circuit:

\begin{theorem}
There exist the following $O(1)$-depth quantum circuits for {\rm
 TH}$_n^t$:
\begin{itemize}
\item An $O(n\log n)$-size circuit for any $1 \leq t \leq \log n$ or
      $n-\log n \leq t \leq n$.

\item An $O(n\sqrt{t\log n})$-size circuit for any $\log n \leq t \leq
      \lceil n/2 \rceil$.

\item An $O(n\sqrt{(n-t)\log n})$-size circuit for any $\lceil n/2
      \rceil \leq t \leq n-\log n$.
\end{itemize}
\end{theorem}
Theorem~2 implies the size difference between the $\qnc$ and $\qtc$
circuits for implementing the same quantum operation. Let $U_n$ be a
quantum operation on $n$ qubits. Let us assume that we have an
optimal-size $\qtc$ circuit for $U_n$ and its size is represented by
some polynomial $s(n)$. Similarly, let $t(n)$ $(\geq s(n))$ be the
optimal $\qnc$ circuit size. The definition of $\qnc$ only implies that
$t(n)$ is bounded above by poly$(n)$. Theorem~2 tells us more about
this: $t(n)$ is $O(s(n)\sqrt{s(n)\log n})$. This is because we can
obtain an $O(s(n)\sqrt{s(n)\log n})$-size $\qnc$ circuit for $U_n$ by
transforming every threshold gate in the optimal-size $\qtc$ circuit
into the $\qnc$ circuit by Theorem~2.

A key ingredient of the circuits in Theorem~2 is an $O(1)$-depth
$O(n^2)$-size quantum circuit for the quantum counting operation, which
computes the counting function on $n$ bits that outputs the binary
representation of the Hamming weight of the input. Our idea for
constructing the circuit is that, after we apply H{\o}yer and {\v 
S}palek's OR reduction, we implement a particular type of the quantum
Fourier transform (QFT) on $O(\log n)$ qubits in depth $O(1)$ and with
size exponential in $\log n$. The QFT part performs many projective
measurements in parallel and applies the circuit in Theorem~1 to the
classical outcomes of the measurements to estimate the phase of a
Fourier state. It is similar to the $O(\log n)$-depth $O(n\log n)$-size
quantum circuit for approximating the QFT on $n$ qubits
\cite{Cleve}. The main difference is that the QFT part requires
exponentially more gates than those in \cite{Cleve} to construct an
$O(1)$-depth exact circuit. Nevertheless, the size is still poly$(n)$
since the input size is $O(\log n)$.

Lastly, we apply Theorem~1 to studying the relationship between $\qnc$
and efficient classical computation. More concretely, based on
Theorem~1, we study the existence of a classically hard
problem\footnote{We deal with not only a decision problem, but also a
relation problem, where a relation problem can have many valid
(polynomial-length) outputs for an input. An algorithm for solving such
a problem outputs any one of them \cite{Aaronson}.} that is  solvable
exactly by a $\qnc$ circuit, where a problem is said to be classically
hard if it cannot be solved by a polynomial-time bounded-error classical
algorithm. To do this, we consider the question of whether a
polynomial-time exact classical algorithm using a $\qnc$ oracle can be
constructed for a discrete logarithm problem (DLP) that seems
classically hard. Here, the $\qnc$ oracle solves, in classical constant
time, a problem that is solvable exactly by a $\qnc$ circuit. Such an
algorithm for the DLP implies the existence of the desired problem under
the plausible assumption that the DLP is classically hard. This is
because the algorithm with a polynomial-time bounded-error classical
simulation of the $\qnc$ oracle would imply that the DLP is not
classically hard.

Based on Shor's bounded-error quantum algorithm for the general 
DLP \cite{Shor}, H{\o}yer and {\v S}palek showed that there exists 
a polynomial-time bounded-error classical algorithm using a 
bounded-error version of the $\qnc$ oracle \cite{Hoyer}. It is, 
however, difficult to directly transform the algorithm into an exact 
one. Based on van Dam's exact quantum algorithm for the general 
DLP~\cite{van}, which is simpler than Mosca and Zalka's \cite{Mosca}, 
we show that, using Theorem~1, under an assumption about the QFT, there
exists the desired algorithm for a particular type of the DLP that seems
classically hard:

\begin{theorem}
Let $q$ be a safe prime, i.e., a prime of the form $2p+1$ for some 
prime $p$, and $n=\lceil \log q\rceil$. If the QFT modulo $p$ is in 
$\qnc$, there exists a poly$(n)$-time exact classical algorithm for 
the DLP over the multiplicative group of integers modulo $q$ using 
the $\qnc$ oracle.
\end{theorem}
We note that, as in the cryptographic literature, we assume that there
exist infinitely many safe~primes. Since we require the assumption about
the QFT, Theorem~3 does not imply the existence of the above-mentioned
problem (under a plausible assumption). It, however, allows us to deepen
our understanding of the relationship among $\qnc$, the QFT, and
efficient classical computation. In fact, it implies that, under the
plausible assumption that the DLP in Theorem 3 is classically hard,
there exists a classically hard problem that is solvable exactly by a
$\qnc$ circuit with gates for the QFT modulo $p$.

Theorem~3 suggests the following key problem for further understanding
the relationship between $\qnc$ and efficient classical or quantum
computation: Is the QFT modulo $p$ in $\qnc$? If this is the case,
Theorem~3 implies the existence of a classically hard problem that is
solvable exactly by a $\qnc$ circuit (under a plausible assumption). If
not, $\qnc$ is strictly weaker than efficient quantum computation, more
precisely, it is strictly contained in the class of quantum operations
implementable approximately (or even exactly) by polynomial-size quantum
circuits. This is because the QFT modulo $p$ is in the latter class
\cite{Mosca,Hales,Hoyer}. We leave the problem about the QFT modulo $p$
as an open problem.

The main components of (a slightly modified version of) van Dam's
algorithm for the DLP are the QFT modulo $p$, arithmetic operations such
as modular exponentiation, and an amplitude amplification procedure
\cite{Brassard}. Our rigorous analysis of the algorithm shows that these
components excluding the QFT can be implemented by using the OR
functions and iterated multiplications with values pre-computed by
polynomial-time exact classical algorithms. This analysis with Theorem 1
implies Theorem~3.

The remainder of this paper is organized as follows. In Section~2, we
give some definitions and the idea of the OR reduction to describe our
results precisely. In Sections~3 and~4, we describe the circuits in
Theorems~1 and 2, respectively. In Section~5, we describe the algorithm
in Theorem 3. In Section~6, we give some open problems. Most of the
proofs are given in Appendix~A.

\section{Preliminaries}

\subsection{Quantum Circuits and Complexity Classes}

We use the standard notation for quantum states and the standard
diagrams for quantum circuits~\cite{Nielsen}. A quantum circuit consists
of elementary gates, where the elementary gates are one-qubit, CNOT, and
unbounded fan-out gates (unless otherwise stated). An unbounded fan-out
gate on $k+1$ qubits implements the quantum operation defined as
$$\left(|y\rangle\bigotimes_{j=0}^{k-1}|x_j\rangle\right)
\mapsto |y\rangle\bigotimes_{j=0}^{k-1}|x_j\oplus y\rangle,$$
where $y,x_j \in \{0,1\}$, $k\geq 1$, and $\oplus$ denotes addition
modulo 2. The first input qubit, i.e., the qubit in state $|y\rangle$,
is called the control qubit. When $k=1$, the gate is a CNOT gate. Since
an unbounded fan-out gate makes copies of a classical source bit, we may
say ``copy'' when we apply this gate. The complexity measures of a
quantum circuit are its size and depth. The size of a quantum circuit is
defined as the total size of all elementary gates in it, where the size
of an elementary gate is defined as the number of qubits affected by the
gate. The depth of a quantum circuit is defined as follows. Input qubits
are considered to have depth 0. For each gate $G$, the depth of $G$ is
equal to 1 plus the maximal depth of a gate on which $G$ depends. The
depth of a quantum circuit is defined as the maximal depth of a gate in
it. Intuitively, the depth is the number of layers in the circuit, where
a layer consists of gates that can be applied in parallel. A quantum
circuit can use ancillary qubits initialized to $|0\rangle$.

For any $a=a_0\cdots a_{n-1}\in \{0,1\}^n \setminus \{0^n\}$, the parity
function with value $a$ on $n$ bits, denoted as PA$_n^a$, is defined as
${\rm PA}_n^a(x)=\bigoplus_{j=0}^{n-1}a_jx_j,$
where $x=x_0\cdots x_{n-1}\in\{0,1\}^n$. We denote PA$_n^{1^n}$ as
PA$_n$. For example, PA$_2^{10}(x)=x_0$, PA$_2^{01}(x)=x_1$, and
PA$_2^{11}(x)={\rm PA}_2(x)=x_0 \oplus x_1$. For any integer $1\leq t
\leq n$, the threshold function with a threshold $t$ on $n$ bits,
denoted as TH$_n^t$, is defined as ${\rm TH}_n^t(x)= 1$ if $|x|\geq t$
and 0 otherwise, where $x=x_0\cdots x_{n-1}\in\{0,1\}^n$ and
$|x|=\sum_{j=0}^{n-1}x_j$, the Hamming weight of $x$. The 
OR function on $n$ bits, denoted as OR$_n$, is defined as TH$_n^1$. The 
AND function on $n$ bits, denoted as AND$_n$, is defined as TH$_n^n$. 
For any integer $1\leq t \leq n$, the exact function with value $t$ on 
$n$ bits, denoted as EX$_n^t$, is defined similarly to TH$_n^t$ except 
that $|x|\geq t$ in the definition of TH$_n^t$ is replaced with 
$|x|=t$. The function EX$_n^0$ is defined as the negation of OR$_n$. 
The quantum operation for computing PA$_n^a$ is defined as
$$\left(\bigotimes_{j=0}^{n-1}|x_j\rangle\right)|z\rangle \mapsto
\left(\bigotimes_{j=0}^{n-1}|x_j\rangle\right)|z\oplus {\rm
PA}_n^a(x)\rangle,$$
where $x_j,z \in \{0,1\}$ and $x=x_0\cdots x_{n-1}$. For simplicity,
this operation is also denoted as PA$_n^a$. The quantum operations
TH$_n^t$, OR$_n$, AND$_n$, and EX$_n^t$ are defined similarly. For any
integer $m > 0$, the quantum Fourier transform modulo $m$, denoted as 
F$_m$, is the quantum operation on  $\lceil \log m \rceil$ qubits 
defined as 
$|x\rangle \mapsto \frac{1}{\sqrt{m}} \sum_{y=0}^{m-1} \omega_m^{xy}
|y\rangle$, where $0 \leq x \leq m-1$ and $\omega_m=e^{2\pi i/m}$.

The quantum complexity class $\qnc$ is the class of quantum
operations implementable exactly by (uniform families of) constant-depth
polynomial-size quantum circuits consisting of the elementary gates
described above. The definition of $\qac$ is the same as that of $\qnc$
except that quantum circuits can use a gate for OR$_k$ as an elementary
gate for any $k$ bounded above by an arbitrary poly$(n)$ for input
length $n$. The definition of $\qtc$ is the same as that of $\qac$
except that quantum circuits can use a gate for TH$_k^t$ as an
elementary gate for any $k$ bounded above by an arbitrary poly$(n)$ and
$1 \leq t \leq k$. Although some authors assume that quantum circuits
can use only a bounded number of distinct one-qubit gates \cite{Hoyer},
we do not assume this since we consider the exact setting. Thus, the
complexity classes in this paper are equal to or larger than those in
the papers that considered only a bounded number of distinct one-qubit
gates. We note, however, that one-qubit gates used in our circuits are
only Hadamard gates $H$ and $Z(\pm\pi/2^k)$ gates for any integer $k\geq
0$, where, for any $\theta \in {\mathbb R}$,
$$H=\frac{1}{\sqrt{2}}
\bigg(
\begin{array}{cc}
1 & 1\\
1 & -1
\end{array}
\bigg),\
Z(\theta)=
\bigg(
\begin{array}{cc}
1 & 0\\
0 & e^{i\theta}
\end{array}
\bigg).
$$

\subsection{H{\o}yer and {\v S}palek's OR Reduction}

\begin{figure}
\vspace{0.2cm}
\centering
\epsfig{file=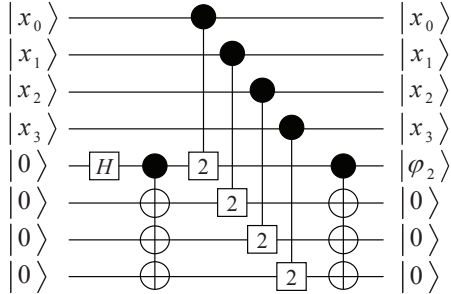,scale=.6}
\vspace{-7.2cm}
\caption{The quantum circuit for preparing $|\varphi_2\rangle$ when
 $n=4$. The gate next to the Hadamard gate is an unbounded fan-out gate
 on four qubits, where the top qubit is the control qubit. The gate
 represented as ``2'' is a $Z(\pi/2^2)$ gate.}
\label{figure1}
\end{figure}

The OR reduction is described as an $O(1)$-depth $O(n\log n)$-size
quantum circuit for exactly reducing the problem of computing OR$_n$ to
that of computing OR$_m$, where $m=\lceil \log(n+1)\rceil$. We explain
the idea of the circuit, which will be used in our circuits. We want to
compute OR$_n$ and let $|x\rangle = |x_0\rangle\cdots|x_{n-1}\rangle$ be
an input state, where $x_j\in\{0,1\}$. The circuit outputs the $m$-qubit
state $\bigotimes_{k=0}^{m-1}H|\varphi_k\rangle$, where
$$|\varphi_k\rangle =
\frac{|0\rangle+e^{i\pi\frac{|x|}{2^k}}|1\rangle}{\sqrt{2}}$$
for any $0\leq k \leq m-1$. If $|x|=|x_0\cdots x_{n-1}|=0$,
$H|\varphi_k\rangle=|0\rangle$ for any $0\leq k \leq m-1$ and thus the
output state is $|0\rangle^{\otimes m}$. If $|x|\geq 1$, there exist
$0\leq a \leq m-1$ and $b\geq 0$ such that $|x|=2^a(2b+1)$. A direct
calculation shows that $H|\varphi_a\rangle=|1\rangle$ and thus the
output state is orthogonal to $|0\rangle^{\otimes m}$. Therefore, the
circuit exactly reduces the problem of computing OR$_n$ to that of
computing OR$_m$. For any $0 \leq k \leq m-1$, $|\varphi_k\rangle$ can
be prepared by an $O(1)$-depth $O(n)$-size quantum circuit as depicted
in Fig.~\ref{figure1}. By using unbounded fan-out gates, all the states
$|\varphi_k\rangle$ can be prepared in parallel and thus the depth and
size of the circuit for the OR reduction are $O(1)$ and $O(nm)=O(n\log
n)$, respectively.

\section{Circuit for the OR Function}

\subsection{Exponential-Size Circuit}

For any Boolean function $f_n:\{0,1\}^n \to \{0,1\}$ satisfying
$f_n(0^n)=0$, there exists a set of real numbers $\{r_a\}_{a\in
\{0,1\}^n \setminus \{0^n\}}$ such that
$$f_n(x)=\sum_{a\in \{0,1\}^n \setminus \{0^n\}}r_a{\rm PA}_n^a(x)$$
for any $x\in \{0,1\}^n$. This is shown by using the Fourier expansion
of $f_n$ \cite{Odonnell}, more precisely, by replacing the Fourier basis
in the Fourier expansion of $f_n$ with a basis consisting of the parity
functions PA$_n^a$. In particular, the following representation of
OR$_n$ can be obtained by using the Fourier expansion of OR$_n$. The
proof is given in Appendix A.1.

\begin{lemma} For any $x\in \{0,1\}^n$, ${\rm
 OR}_n(x)=\frac{1}{2^{n-1}}\sum_{a\in \{0,1\}^n \setminus \{0^n\}}{\rm
 PA}_n^a(x)$.
\end{lemma}

\begin{figure}[t]
\vspace{0.3cm}
\centering
\epsfig{file=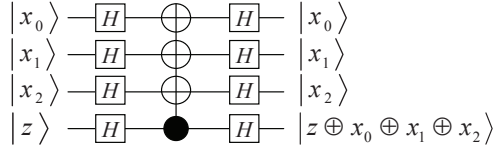,scale=.6}
 \vspace{-9.2cm}
\caption{The quantum circuit for PA$_3$ \cite{Green}.}
\label{figure2}
\end{figure}

The representation of OR$_n$ implies an $O(1)$-depth $O(n2^n)$-size
quantum circuit for OR$_n$. The idea is that, when the input $x$ is
given, we compute PA$_n^a(x)$ for every $a$ in parallel and prepare the
state $(|0\rangle^{\otimes (2^n-1)} + (-1)^{{\rm OR}_n(x)}
|1\rangle^{\otimes (2^n-1)})/\sqrt{2}$ based on the
representation. Applying an unbounded fan-out gate and a Hadamard gate
to the state gives the desired state $|{\rm OR}_n(x)\rangle$. The point
is that there exists an $O(1)$-depth $O(|a|)$-size quantum circuit for
PA$_n^a$ consisting of Hadamard gates and an unbounded fan-out gate as
depicted in Fig.~\ref{figure2}~\cite{Green}.

To describe the circuit for OR$_n$ more precisely, let
$|x\rangle=|x_0\rangle\cdots |x_{n-1}\rangle$ be an input state. The
circuit is described as follows:
\begin{enumerate}
\item Copy the input state $|x\rangle$ and apply the circuit for
      PA$_n^a$ to each copy for every $a\in \{0,1\}^n \setminus
      \{0^n\}$ in parallel to prepare the state $\bigotimes_{a\in
      \{0,1\}^n \setminus \{0^n\}}|{\rm PA}_n^a(x)\rangle$.

\item Apply a Hadamard gate and an unbounded fan-out gate to ancillary
      qubits (initialized to $|0\rangle$) to prepare the $(2^n-1)$-qubit
      state $(|0\rangle^{\otimes (2^n-1)}+ |1\rangle^{\otimes
      (2^n-1)})/\sqrt{2}.$

\item Apply controlled-$Z(\pi/2^{n-1})$ gates in parallel to the states
      in Steps~1 and 2 to prepare the state
$$\frac{|0\rangle^{\otimes (2^n-1)}+ e^{i\pi
 \frac{1}{2^{n-1}}\sum_{a\in \{0,1\}^n \setminus \{0^n\}}{\rm
 PA}_n^a(x)}|1\rangle^{\otimes (2^n-1)}}{\sqrt{2}}
= \frac{|0\rangle^{\otimes (2^n-1)} + (-1)^{{\rm OR}_n(x)}
|1\rangle^{\otimes (2^n-1)}}{\sqrt{2}},$$
where Lemma~1 implies the equation.

\item Apply an unbounded fan-out gate and a Hadamard gate to the state
      in Step~3 to prepare the desired state $|{\rm OR}_n(x)\rangle$.
\end{enumerate}

For any $0 \leq j \leq n-1$, let $e(j)=e_0\cdots e_{n-1}\in \{0,1\}^n$
such that $e_k=1$ if $k=j$ and 0 otherwise. In Step~1, since the input
state $|x_j\rangle = |{\rm PA}_n^{e(j)}(x)\rangle$, it suffices to
prepare the state $|{\rm PA}_n^a(x)\rangle$ for every $a\in \{0,1\}^n$
such that $|a| \geq 2$. To prepare the states in parallel, we require
the state $|x_j\rangle^{\otimes (2^{n-1}-1)}$ for any $0 \leq j \leq
n-1$. Thus, before applying the circuit for PA$_n^a$, we apply an
unbounded fan-out gate to the input qubit in state $|x_j\rangle$ and
$2^{n-1}-1$ ancillary qubits for every $0 \leq j \leq n-1$ in
parallel. In Step~2, we apply an unbounded fan-out gate to the ancillary
qubits in state $(H|0\rangle)|0\rangle^{\otimes (2^n-2)}$. In Step~3, we
use the qubit in state $|{\rm PA}_n^a(x)\rangle$ as the control qubit of
the controlled-$Z(\pi/2^{n-1})$ gate. In Step~4, we first apply an
unbounded fan-out gate to the state in Step~3 to disentangle the last
$2^n-2$ qubits and obtain the state $(|0\rangle + (-1)^{{\rm
OR}_n(x)}|1\rangle)/\sqrt{2}$. Thus, the Hadamard gate outputs the
desired state. By the construction, the depth of the whole circuit does
not depend on $n$. Since Step~1 is the dominant part and uses $n$
unbounded fan-out gates on $2^{n-1}$ qubits, the size of the whole
circuit is $O(n2^n)$. This implies the following lemma. The details of
the proof are given in Appendix A.2.

\begin{lemma} There exists an $O(1)$-depth $O(n2^n)$-size quantum circuit
 for {\rm OR}$_n$.
\end{lemma}
{\bf Remark:} Hoban et al.\ considered a restricted model of
 measurement-based quantum computation, where the adaptivity of
 measurements is removed \cite{Hoban}. They showed that, if we are
 allowed to use the $(2^n-1)$-qubit state in Step~2, any Boolean
 function $f_n$ can be computed exactly in the model by the procedure
 based on the above-mentioned representation of $f_n$. The circuit in
 Lemma~2 can be considered as a simulation of the procedure for
 computing OR$_n$ in the model. The unbounded fan-out gates are mainly
 used for preparing the $(2^n-1)$-qubit state and for computing
 PA$_n^a$.

\subsection{Proof of Theorem~1}

We show Theorem~1 using H{\o}yer and {\v S}palek's OR reduction and
Lemma~2. Let $|x\rangle=|x_0\rangle\cdots |x_{n-1}\rangle$ be an input
state. The circuit is described as follows:
\begin{enumerate}
\item Apply H{\o}yer and {\v S}palek's OR reduction to the input state
      $|x\rangle$ to prepare the $m$-qubit state
      $\bigotimes_{k=0}^{m-1}H|\varphi_k\rangle$, where $m=\lceil
      \log(n+1)\rceil$.

\item Apply the circuit in Lemma~2 to the state in Step~1 to prepare the
      desired state $|{\rm OR}_n(x)\rangle$.
\end{enumerate}
Since Step~1 exactly reduces the problem of computing OR$_n$ to that of
computing OR$_m$ in depth $O(1)$ and with size $O(n\log n)$, Step~2
outputs the desired state. Since the input size to Step~2 is $m$, the
depth and size of the circuit in Step~2 are $O(1)$ and $O(m2^m)=O(n\log
n)$, respectively. Thus, the depth and size of the whole circuit are
$O(1)$ and $O(n\log n)$, respectively. This completes the proof.

Theorem~1 immediately implies that OR$_n$ is in $\qnc$ and thus the 
following relationship holds:

\begin{corollary}
$\qnc = \qac$.
\end{corollary}
Since $\qac = \qtc$ \cite{Hoyer}, it holds that $\qnc = \qac = \qtc$. 
Corollary~1 and the relationship $\qac = \qtc$ immediately imply that
$\qnck=\qack$ and $\qack = \qtck$, respectively, for any integer 
$k \geq 0$, where $\qnck$, $\qack$, and $\qtck$ are defined similarly to
$\qnc$, $\qac$, and $\qtc$, respectively, except that they deal with 
$O(\log^k n)$-depth circuits in place of $O(1)$-depth ones. Therefore, 
more generally, it holds that $\qnck = \qack = \qtck$ for any integer 
$k \geq 0$.

For any integer constant $c\geq 1$, the size of the circuit in Theorem~1
can be decreased to $O(n\log^{(c)}n)$ without increasing the depth
asymptotically, where $\log^{(c)}n$ is the $c$-times iterated logarithm
$\log\cdots\log n$. To show this, we divide the $n$ input qubits 
into $n/\log n$
blocks of $\log n$ qubits. For each block, we apply the circuit in
Theorem~1 to compute OR$_{\log n}$. We obtain $n/\log n$ output qubits
and apply the circuit again to the output qubits to compute OR$_{n/\log
n}$, which yields the desired output. The depth and size of the whole
circuit are $O(1)$ and $O(n\log^{(2)}n)$, respectively. Using the
resulting circuit, we repeat this size-reduction procedure. After $c-1$
times repetition, we obtain an $O(n\log^{(c)}n)$-size circuit.

The circuit for OR$_n$ yields a circuit for EX$_n^t$ \cite{Hoyer}. To
construct the circuit, it suffices to prepare
$Z(-t\pi/2^k)|\varphi_k\rangle$ in place of $|\varphi_k\rangle$ in
H{\o}yer and {\v S}palek's OR reduction and to negate the final output
of the circuit in Theorem~1. This is done by only adding a
$Z(-t\pi/2^k)$ gate for every $0 \leq k \leq m-1$ and a NOT gate. Thus,
the depth and size of the resulting circuit are asymptotically the same
as those in Theorem~1. This yields an $O(1)$-depth $O(n\log n)$-size
quantum circuit for EX$_n^t$ for any $0 \leq t \leq n$.

\section{Circuit for the Threshold Function}

First, we describe a constant-depth circuit for TH$_n^t$ based on the
constant-depth circuits for EX$_n^k$ described above. Then, we describe
another constant-depth circuit for TH$_n^t$ based on a circuit for the
counting function. Next, we combine these two circuits to show
Theorem~2.

\subsection{Exact-Function-Based and Counting-Function-Based Circuits}

We first consider a constant-depth circuit for TH$_n^t$ based on the
circuits for EX$_n^k$ when $1 \leq t \leq \lceil n/2 \rceil$. Let
$|x\rangle = |x_0\rangle\cdots|x_{n-1}\rangle$ be an input state. The
circuit is described as follows:
\begin{enumerate}
\item Copy the input state $|x\rangle$ and apply the circuit for
      EX$_n^k$ to each copy for every $0\leq k \leq t-1$ in parallel to
      prepare the state $\bigotimes_{k=0}^{t-1}|{\rm
      EX}_n^k(x)\rangle$.

\item Apply the circuit for PA$_t$ and a NOT gate to the state in Step~1
      to prepare the state $|\bigoplus_{k=0}^{t-1} {\rm EX}_n^k(x)
      \oplus 1\rangle$.
\end{enumerate}

If $|x|\geq t$, EX$_n^k(x)=0$ for every $0 \leq k \leq t-1$. If $|x|<t$,
there exists exactly one $0 \leq k \leq t-1$ such that
EX$_n^k(x)=1$. Thus, the state in Step~2 is equal to the desired state
$|{\rm TH}_n^t(x)\rangle$. The depth and size of the circuit in Step~1
are $O(1)$ and $O(tn\log n)$, respectively. As depicted in
Fig.~\ref{figure2}, the depth and size of the circuit for PA$_t$ are
$O(1)$ and $O(t)$, respectively. Thus, the depth and size of the whole
circuit are $O(1)$ and $O(tn\log n)$, respectively. When $\lceil n/2
\rceil \leq t \leq n$, we modify the circuit in such a way that it
prepares the state $|\bigoplus_{k=t}^n {\rm EX}_n^k(x)\rangle$ in
Step~2. This implies the following lemma:

\begin{lemma} There exist the following $O(1)$-depth quantum circuits
 for {\rm TH}$_n^t$:
\begin{itemize}
\item An $O(tn\log n)$-size circuit for any $1 \leq t \leq
      \lceil n/2 \rceil$.

\item An $O((n-t+1)n\log n)$-size circuit for any $\lceil n/2 \rceil
      \leq t \leq n$.
\end{itemize}
\end{lemma}
When $t$ is an integer constant, the size is $O(n\log n)$. On the other
hand, when $t=\lceil n/2 \rceil$, in other words, for the majority
function, the size is $O(n^2\log n)$.

We define the counting function on $n$ bits, denoted as CO$_n$, as ${\rm
CO}_n(x)=s_0\cdots s_{m-1},$ where $x\in\{0,1\}^n$, $s_j\in\{0,1\}$,
$m=\lceil \log(n+1)\rceil$, and $|x|=\sum_{j=0}^{m-1}s_j2^j$. It
computes the binary representation of the Hamming weight of the
input. The quantum operation for computing CO$_n$ is defined as
$$\left(\bigotimes_{j=0}^{n-1}|x_j\rangle\right)
\left(\bigotimes_{j=0}^{m-1}|z_j\rangle\right) \mapsto
\left(\bigotimes_{j=0}^{n-1}|x_j\rangle\right)
\left(\bigotimes_{j=0}^{m-1}|z_j\oplus s_j\rangle\right),$$
where $x_j,z_j \in \{0,1\}$. This operation is also denoted as
CO$_n$.

We construct a constant-depth circuit for CO$_n$. Let $|x\rangle$ be an 
input state. Since $|x|=\sum_{j=0}^{m-1}s_j2^j$, $|\varphi_k \rangle$ 
in H{\o}yer and {\v S}palek's OR reduction is
$(|0\rangle + e^{i\pi \sum_{j=0}^k \frac{s_j} {2^{k-j}}} |1\rangle )/
\sqrt{2}$. This implies that
$|\varphi_0\rangle\cdots |\varphi_{m-1}\rangle = {\rm F}_{2^m}
|s_0\rangle \cdots |s_{m-1}\rangle$. Thus, to obtain the desired state
$|s_0\rangle\cdots |s_{m-1}\rangle$, it suffices to implement the
following type of the inverse of the QFT: $|x\rangle ({\rm
F}_{2^m}|s_0\rangle \cdots |s_{m-1}\rangle) \mapsto |x\rangle
|s_0\rangle \cdots |s_{m-1}\rangle$. Our idea for implementing this
operation is to perform $A(\theta)$-measurements on many
$|\varphi_k\rangle$'s in parallel for appropriate $\theta$'s, where, 
for any $\theta \in {\mathbb R}$, an $A(\theta)$-measurement is the
one-qubit projective measurement in the basis $(|0\rangle + 
e^{i\theta}|1\rangle)/\sqrt{2}, (|0\rangle - 
e^{i\theta}|1\rangle)/\sqrt{2}$, which correspond to the classical 
outcomes 0 and~1, respectively. The classical outcomes imply each 
$s_k$ exactly. 

For example, when $m=3$, we first prepare the state $|\varphi_0\rangle
|\varphi_1\rangle^{\otimes 2} |\varphi_2\rangle^{\otimes 4}$ with a
slightly modified version of H{\o}yer and {\v S}palek's OR reduction,
where
$$|\varphi_0 \rangle = \frac{|0\rangle + e^{i\pi s_0} |1\rangle
}{\sqrt{2}},
|\varphi_1 \rangle = \frac{|0\rangle + e^{i\pi (s_1 + \frac{s_0}{2})}
|1\rangle}{\sqrt{2}},
|\varphi_2 \rangle = \frac{|0\rangle + e^{i\pi (s_2 + \frac{s_1}{2} +
\frac{s_0}{2^2})} |1\rangle}{\sqrt{2}}.$$
We can easily obtain $s_0$ since it is equal to the classical outcome
$s_0^\varepsilon$ of an $A(0)$-measurement on $|\varphi_0\rangle$. The
value $s_1$ is determined depending on $s_0$. When $s_0=0$, $s_1$ is
equal to the classical outcome $s_1^0$ of an $A(0)$-measurement on
$|\varphi_1\rangle$. When $s_0=1$, $s_1$ is equal to the classical
outcome $s_1^1$ of an $A(\pi/2)$-measurement on $|\varphi_1\rangle$. In
other words, $s_1=s_1^{s_0}$. Similarly, we perform $A(0)$-,
$A(\pi/4)$-, $A(\pi/2)$-, and $A(3\pi/4)$-measurements on
$|\varphi_2\rangle^{\otimes 4}$ and let $s_2^{00}$, $s_2^{10}$,
$s_2^{01}$, and $s_2^{11}$ be the classical outcomes, respectively. By
the definition of the measurements, $s_2=s_2^{s_0s_1}$. These
relationships imply
\begin{eqnarray*}
s_1 &=& [s_1^0(1\oplus 0 \oplus s_0^\varepsilon)]\oplus[s_1^1(1\oplus
 1 \oplus s_0^\varepsilon)],\\
s_2 &=& [s_2^{00}(1\oplus 0 \oplus s_0^\varepsilon)(1\oplus 0 \oplus
 s_1^0)]\oplus [s_2^{10}(1\oplus 1 \oplus s_0^\varepsilon)(1\oplus 0 
 \oplus s_1^1)]\\
&& \oplus [s_2^{01}(1\oplus 0 \oplus s_0^\varepsilon)(1\oplus 1 \oplus
 s_1^0)]\oplus [s_2^{11}(1\oplus 1 \oplus s_0^\varepsilon)(1\oplus 1
 \oplus s_1^1)].
\end{eqnarray*}
Thus, if we have sufficiently many copies of the classical outcomes, we
can compute $s_k$ for every $1 \leq k \leq m-1$ in parallel using the
circuits for AND$_{k+1}$ and PA$_{2^k}$. We note that we can perform all
the above measurements in parallel. We define the function $t_k(y)$ on
$k$ bits as $t_k(y) = s_k^y\bigwedge_{j=0}^{k-1}(1\oplus y_j\oplus
s_j^{y_0\cdots y_{j-1}})$ for any $y=y_0\cdots y_{k-1}\in\{0,1\}^k$,
where the value $s_j^{y_0\cdots y_{j-1}}$ is regarded as
$s_0^\varepsilon$ when $j=0$. It holds that $s_1=t_1(0)\oplus t_1(1)$
and $s_2=t_2(00) \oplus t_2(10) \oplus t_2(01) \oplus t_2(11)$.

To describe the circuit for CO$_n$ more precisely and generally, let
$|x\rangle=|x_0\rangle\cdots |x_{n-1}\rangle$ be an input state. The
circuit is described as follows:
\begin{enumerate}
\item Apply a slightly modified version of H{\o}yer and {\v S}palek's OR
      reduction to the input state $|x\rangle$ to prepare the state
      $\bigotimes_{k=0}^{m-1}|\varphi_k\rangle^{\otimes 2^k}$.

\item Perform $A(0)$- and
      $A(\pi\sum_{j=0}^{k-1}\frac{y_j}{2^{k-j}})$-measurements for every
      $1 \leq k \leq m-1$ and $y=y_0\cdots y_{k-1}\in \{0,1\}^k$ in
      parallel on the state in Step~1 to obtain the values
      $s_0^\varepsilon,s_k^y \in \{0,1\}$ such that
      $s_0=s_0^\varepsilon$ and $s_k=s_k^{s_0\cdots s_{k-1}}$.
  
\item Prepare $2^m-1$ copies of the state $|s_0^\varepsilon\rangle$ and
      $2^{m-k}-1$ copies of the state $|s_k^y\rangle$ and apply the
      circuit for AND$_{k+1}$ (constructed by the circuit for OR$_{k+1}$
      in Section~3) to the states for every $1 \leq k \leq m-1$ and $y
      \in \{0,1\}^k$ in parallel to prepare the state $\bigotimes_{1
      \leq k \leq m-1,y\in \{0,1\}^k}|t_k(y)\rangle$.

\item Apply the circuit for PA$_{2^k}$ for every $1 \leq k \leq m-1$ in
      parallel to the state in Step~3 to prepare the state
      $|s_0^\varepsilon\rangle\bigotimes_{1 \leq k \leq
      m-1}|\bigoplus_{y\in\{0,1\}^k}t_k(y)\rangle.$
\end{enumerate}

Since $t_k(y)=s_k$ if $y=s_0\cdots s_{k-1}$ and 0 otherwise for any $1
\leq k \leq m-1$, $\bigoplus_{y\in\{0,1\}^k}t_k(y)=s_k$. Thus,
Step~4 outputs the desired state. By the construction, the depth of the
whole circuit does not depend on $n$. Since Step~1 is the dominant part
and the state in Step~1 can be prepared with a circuit of size
$O(n\sum_{k=0}^{m-1}2^k)=O(n^2)$, the size of the whole circuit is
$O(n^2)$. This implies the following lemma. The details of the proof are
given in Appendix A.3.

\begin{lemma} There exists an $O(1)$-depth $O(n^2)$-size quantum circuit
 for {\rm CO}$_n$.
\end{lemma}

Lemma~4 yields an $O(1)$-depth $O(n^2)$-size quantum circuit for
TH$_n^t$. To construct the circuit, it suffices to add a circuit for
comparing $t$ with the output of the circuit for CO$_n$. We can
construct an $O(1)$-depth ${\rm poly}(m)$-size quantum circuit for the
comparison using the circuit for addition in \cite{Chandra}.

\subsection{Combination of the Two Circuits}

A careful combination of the circuits in Lemmas~3 and 4 yields a
smaller circuit for TH$_n^t$. We explain the idea in the case when $1
\leq t \leq \lceil n/2 \rceil$. When the input $x$ is given, before
using the first circuit in Lemma~3, we compute some low-order bits (not
all the bits!) of the binary representation of $|x|$ by the circuit in
Lemma~4. Since we know the low-order bits, it is not necessary to check
whether EX$_n^k(x)=1$ for every $0 \leq k \leq t-1$ as in Lemma~3. It
suffices to consider $0 \leq k \leq t-1$ such that the low-order bits of
the binary representation of $k$ are equal to those computed by the
circuit in Lemma~4. The number of $k$'s we need to consider is decreased
and thus the size of the whole circuit can be decreased.

More precisely, the circuit is described as follows:
\begin{enumerate}
\item  Apply the circuit in Lemma~4 to the input state $|x\rangle =
       |x_0\rangle\cdots|x_{n-1}\rangle$ to prepare the state
       $|s_0\rangle \cdots |s_{l-1}\rangle$, where $s_0\cdots s_{l-1}$
       are the $l$ low-order bits of the binary representation of $|x|$
       and $l$ is an integer satisfying $0 \leq l < \lceil \log (t+1)
       \rceil$.

\item  Apply the first circuit in Lemma~3 to the input state
       $|x\rangle$ to prepare the state $|\bigoplus_k {\rm EX}_n^k(x)
       \oplus 1\rangle$, where we consider only $0 \leq k \leq t-1$ such
       that the $l$ low-order bits of the binary representation of $k$
       are equal to $s_0\cdots s_{l-1}$.
\end{enumerate}

Step~2 outputs the desired state as in Lemma~3. It is obvious that the
depth does not depend on~$n$. The size of the circuit in Step~1 is
$O(2^ln)$ and that in Step~2 is $O(2^{-l}tn\log n)$ since there are at
most $2^{-l}t$ $k$'s we need to consider. The same idea with the second
circuit in Lemma~3 works when $\lceil n/2 \rceil \leq t \leq n$. This
implies the following lemma. The details of the proof are given in
Appendix A.4.

\begin{lemma} There exist the following $O(1)$-depth quantum circuits
 for {\rm TH}$_n^t$:
\begin{itemize}
\item An $O(2^ln+2^{-l}tn\log n)$-size circuit for any $1
      \leq t \leq \lceil n/2 \rceil$ and $0 \leq l < \lceil \log (t+1)
      \rceil$.

\item An $O(2^ln+2^{-l}(n-t+1)n\log n+n\log n)$-size circuit for any
      $\lceil n/2 \rceil \leq t \leq n$ and $0 \leq l < \lceil \log
      (t+1) \rceil$.
\end{itemize}
\end{lemma}

By setting $l$ appropriately depending on $t$, Lemma~5 implies
Theorem~2. The proof is given in Appendix A.5. The size of the circuit
for TH$_n^{\lceil n/2 \rceil}$ in Lemma~3 is $O(n^2\log n)$ and it can
be decreased to $O(n^2)$ by Lemma~4. Theorem~2 with $t= \lceil
n/2\rceil$ yields an even smaller circuit:

\begin{corollary}
There exists an $O(1)$-depth $O(n\sqrt{n\log n})$-size quantum circuit
 for {\rm TH}$_n^{\lceil n/2\rceil}$.
\end{corollary}

\section{Discrete Logarithm Algorithm Using a $\qnc$ Oracle}

Let $q > 5$ be a safe prime, i.e., a prime of the form $q=2p+1$ for some
prime $p > 2$. In the following, as in the cryptographic literature, we
assume that there exist infinitely many safe primes. Let $G_q = ({\mathbb
Z}/q{\mathbb Z})^*$, the multiplicative group of integers modulo $q$. It
is known that there exists a generator $1 < g_q \leq q-1$ of $G_q$ and
thus $G_q= \{g_q^0=1,g_q^1,\ldots,g_q^{q-2}\}$ and $g_q^{q-1} \equiv 1 \
{\rm mod} \ q$. The discrete logarithm problem (DLP) over $G_q$ (with
respect to given $q$ and $g_q$) is to find $0 \leq l_q \leq q-2$ such
that $g_q^{l_q} \equiv x_q \ {\rm mod} \ q$ for an input $x_q \in G_q$,
where the problem size is $n=\lceil \log q \rceil$ and the order of
$G_q$, i.e., $q-1$ and its decomposition $2p$ are known. Since it seems
difficult to reduce the DLP over $G_q$ to DLP's over groups of
sufficiently small orders, it is plausible that it cannot be solved by a
polynomial-time bounded-error classical algorithm, in other words, that
the DLP over $G_q$ is classically hard.

Although we can directly consider the DLP over $G_q$, for simplicity, 
we consider simpler DLP's obtained by the reduction method in
\cite{Pohlig}. Since the order of $G_q$ is $2p$ and $\gcd(2,p)=1$, the
DLP over $G_q$ with an input $x_q$ can be reduced to the following two
DLP's by a poly$(n)$-time exact classical algorithm. One is the DLP over
the group generated by $g_q^p$ with the input $x_q^p$, which is solvable
by a poly$(n)$-time exact classical algorithm since the order of $g_q^p$
is 2. The other is the DLP over the group $G$ generated by $g=g_q^2$
with the input $x=x_q^2$. Thus, to show Theorem 3, it suffices to show
that, if F$_p$ is in $\qnc$, there exists a poly$(n)$-time exact
classical algorithm for the DLP over $G$ using the $\qnc$ oracle, which
solves, in classical constant time, a problem that is solvable exactly
by a $\qnc$ circuit.

We analyze (a slightly modified version of) van Dam's exact algorithm for
the DLP~\cite{van}, which consists of two parts. The first part is
independent of the input $x \in G$ and transforms the state
$|0\rangle^{\otimes (m+n)}$ into the state $\frac{1}{\sqrt{p-1}}
\sum_{s=1}^{p-1}|s\rangle |\chi^s\rangle$ as follows, where $m=\lceil
\log p \rceil$, the $n$-qubit state $|\chi^s\rangle = \frac{1}{\sqrt{p}}
\sum_{r=0}^{p-1} \omega_p^{sr} |g^r \ {\rm mod} \ q\rangle$ for any $0
\leq s \leq p-1$, and $\omega_p=e^{2\pi i/p}$:
\begin{enumerate}
\item Apply F$_p$ to the first $m$ qubits of the state 
$|0\rangle^{\otimes (m+n)}$ to prepare the state
      $\frac{1}{\sqrt{p}} \sum_{r=0}^{p-1} |r\rangle|0\rangle^{\otimes
      n}$.

\item Apply the modular exponentiation operation $|r\rangle|0\rangle \to
      |r \rangle |g^r \ {\rm mod} \ q \rangle$ to the state in Step 1
      to prepare the state $\frac{1}{\sqrt{p}}
      \sum_{r=0}^{p-1}|r\rangle|g^r \ {\rm mod} \ q\rangle$.

\item Apply F$_p$ to the first $m$ qubits of the state in Step 2 to
      prepare the state $\frac{1}{\sqrt{p}} \sum_{s=0}^{p-1}
      |s\rangle|\chi^s\rangle$.

\item Apply the amplitude amplification procedure to prepare the state
      $\frac{1}{\sqrt{p-1}} \sum_{s=1}^{p-1}|s\rangle |\chi^s\rangle$.
\end{enumerate}

Steps 1 and 3 are in $\qnc$ by our assumption. Since $g^r
\equiv \prod_{j=0}^{m-1} g^{2^jr_j} \ {\rm mod} \ q$ when
$r=\sum_{j=0}^{m-1}2^jr_j$ and $r_j \in\{0,1\}$, the modular
exponentiation operation in Step 2 can be implemented by using the
iterated multiplication operation with the values $g^{2^j} \ {\rm mod} \
q$ that can be pre-computed by a poly$(n)$-time exact classical  
algorithm~\cite{Cleve}. It holds that $\qnc = \qtc$ as shown in
Section~3 and $\qtc$ includes arithmetic operations\footnote{To show
this, we need to show that the ``weighted'' threshold gates are in
$\qtc$. We can simply show this as in \cite{Hoyer}.} such as the
iterated multiplication operation \cite{Siu}. Thus, Step 2 is in
$\qnc$.

The procedure in Step 4 is similar to the one in
\cite{Brassard0}. We define the algorithm $\cal A'$ as Steps 1, 
2, and~3, and the good state $|A'\rangle = \frac{1}{\sqrt{p}}
\sum_{s=1}^{p-1}|s\rangle |\chi^s\rangle$. Since $\langle A'|A'\rangle
= 1-1/p$, it is easy to transform $\cal A'$ into a new algorithm $\cal
A$ with success probability 1/2 using one ancillary qubit. Thus, we 
require only one application of a Grover iteration with $\cal A$. The
Grover iteration includes operations that change the phases of the
states $|0\rangle^{\otimes k}$ with some $k \leq m+n+1$. These
operations can be implemented by using OR$_n$, which is in $\qnc$ as
shown in Section 3. Thus, the whole procedure in Step 4 is in $\qnc$.

For the input $x = x_q^2 \equiv g^l \in G$ $(0 \leq l \leq p-1)$, the 
second part of van Dam's exact algorithm transforms the state
$\frac{1}{\sqrt{p-1}} \sum_{s=1}^{p-1}|s\rangle |\chi^s\rangle 
|0\rangle^{\otimes m}$ into the state $\frac{1}{\sqrt{p-1}}
\sum_{s=1}^{p-1}|s\rangle |\chi^s\rangle |sl
\ {\rm mod} \ p\rangle$ as follows:
\begin{enumerate}
\item[5.] Apply F$_p$ to the last $m$ qubits to prepare the state
	  $\frac{1}{\sqrt{p-1}} \sum_{s=1}^{p-1}|s\rangle
	  |\chi^s\rangle\left(\frac{1}{\sqrt{p}} \sum_{\alpha=0}^{p-1}
	  |\alpha\rangle\right)$.

\item[6.] Apply $D_x:|y\rangle|\alpha\rangle \mapsto |y\cdot x^{-\alpha}
	  \ {\rm mod} \ q\rangle|\alpha\rangle$ $(0 \leq y \leq
	  q-1,\ 0 \leq \alpha \leq p-1)$ to the last $n+m$ qubits of
	  the state in Step~5 to prepare the state $\frac{1}{\sqrt{p-1}}
	  \sum_{s=1}^{p-1}|s\rangle |\chi^s\rangle
	  \left(\frac{1}{\sqrt{p}} \sum_{\alpha=0}^{p-1}
	  \omega_p^{sl\alpha}|\alpha\rangle\right)$. Note that
	  $D_x|\chi^s\rangle|\alpha\rangle = \omega_p^{sl\alpha}
	  |\chi^s\rangle |\alpha\rangle$.

\item[7.] Apply F$_p^{-1}$ (as in Step~5) to prepare the state
	  $\frac{1}{\sqrt{p-1}} \sum_{s=1}^{p-1}|s\rangle |\chi^s\rangle
	  |sl \ {\rm mod} \ p\rangle$.
\end{enumerate}
One-qubit projective measurements in the basis $|0\rangle,|1\rangle$ on
the state in Step~7 yield the classical outcomes $s$ and $sl \ {\rm mod}
\ p$ for some $1 \leq s \leq p-1$. Since $\gcd(s,p)=1$, we can 
compute $sl\cdot s^{-1} \ {\rm mod} \ p = l$, which is the desired result,
by a poly$(n)$-time exact classical algorithm. Steps 5 and 7 are in 
$\qnc$ by our assumption. Step 6 is in $\qnc$ since, as in Step~2,  
$D_x$ can be implemented by using arithmetic operations with the 
pre-computed values $x^{2^j} \ {\rm mod} \ q$ and 
$(x^{-1})^{2^j} \ {\rm mod} \ q$. This analysis implies Theorem~3. 
The details of the proof are given in Appendix~A.6.

As described above, the (relation) problem of finding $s$ and $sl \ {\rm
mod} \ p$ for some $1 \leq s \leq p-1$ for the input $x \equiv g^l \in
G$ with the pre-computed values can be solved exactly by the $\qnc$
circuit with gates for F$_p$. On the other hand, the problem is
classically hard under the plausible assumption that the DLP over $G_q$
is classically hard, since otherwise we can easily show that the
plausible assumption does not hold. Thus, under the plausible
assumption, there exists a classically hard problem that is solvable
exactly by a $\qnc$ circuit with gates for F$_p$.

\section{Open Problems}

Interesting challenges would be to find ways of improving our quantum
circuits and to further study the relationships between the
complexity classes. We give some examples of such problems:

\begin{itemize}
\item Does there exist an $O(1)$-depth $O(n)$-size exact or approximate 
quantum circuit for OR$_n$?

\item Does there exist an $O(1)$-depth $O(n\log n)$-size exact 
      quantum circuit for TH$_n^t$ for any $1 \leq t \leq n$?

\item Does it hold that F$_p$ is in $\qnc$?

\item The classes ${\sf QAC^{\rm 0}}$ and ${\sf QTC^{\rm 0}}$ are
      defined similarly to $\qac$ and $\qtc$, respectively, except that
      unbounded fan-out gates are not allowed. Does it hold that ${\sf
      QAC^{\rm 0}} \subsetneq \qac$ or ${\sf QTC^{\rm 0}} \subsetneq
      \qtc$?

\item Does there exist a fundamental gate that is as powerful as an
      unbounded fan-out gate?
\end{itemize}

\appendix

\section{Proofs}

\subsection{Proof of Lemma 1}

We show this lemma by induction on $n$ (without using the Fourier
expansion of OR$_n$ explicitly). It is obvious that the lemma holds when
$n=1$. We assume that it holds when $n=k$. For any $x=x_0\cdots
x_k\in\{0,1\}^{k+1}$,
\begin{eqnarray*}
\frac{1}{2^k}\sum_{a\in \{0,1\}^{k+1} \setminus \{0^{k+1}\}}{\rm
PA}_{k+1}^a(x)
& = & \frac{1}{2^k} \sum_{a\in \{0,1\}^k \setminus \{0^k\}}{\rm
PA}_k^a(x_0\cdots x_{k-1})\\
&& + \frac{x_k}{2^k} +  \frac{1}{2^k}\sum_{a\in \{0,1\}^k \setminus
 \{0^k\}}({\rm PA}_k^a(x_0\cdots x_{k-1})\oplus x_k)\\
& = & 
\left\{
\begin{array}{cc}
{\rm OR}_k(x_0\cdots x_{k-1}), & \mbox{if $x_k=0$,}\\
1, & \mbox{otherwise,}
\end{array}
\right.
\end{eqnarray*}
where the induction hypothesis implies the second equation. The value is
equal to ${\rm OR}_{k+1}(x_0\cdots x_k)$. Thus, when $n=k+1$, the lemma
holds as desired.

\subsection{Proof of Lemma 2}

\begin{figure}[t]
 \vspace{0.2cm}
\centering
\epsfig{file=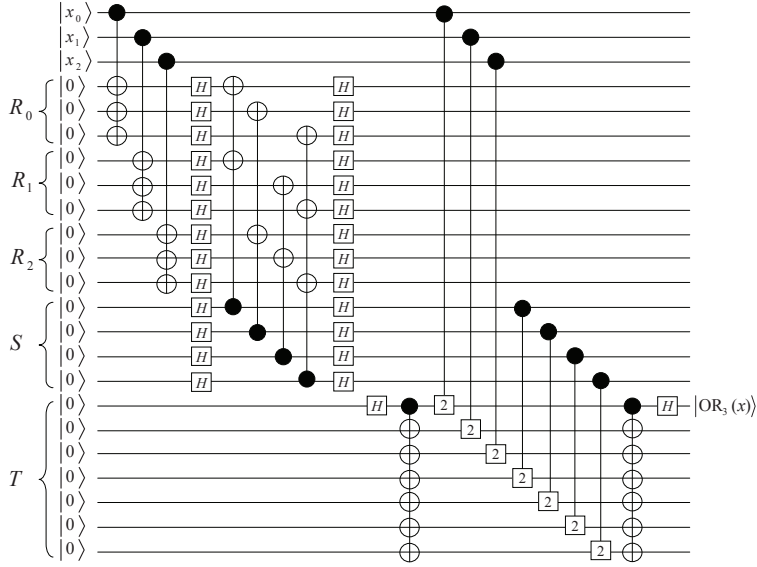,scale=.4}
\vspace{.1cm}
\caption{The circuit for OR$_3$.}
\label{figure3}
\end{figure}

Let $|x\rangle=|x_0\rangle\cdots |x_{n-1}\rangle$ be an input state. As
described in Section~3.1, we prepare the states $|x_j\rangle^{\otimes
(2^{n-1}-1)}$ for any $0\leq j \leq n-1$, $|{\rm PA}_n^a(x)\rangle$ for
any $a\in \{0,1\}^n$ such that $|a| \geq 2$, and the $(2^n-1)$-qubit
state
$$\frac{|0\rangle^{\otimes (2^n-1)} + |1\rangle^{\otimes
(2^n-1)}}{\sqrt{2}}.$$ Thus, we prepare the registers $R_j$ for storing
the state $|x_j\rangle^{\otimes (2^{n-1}-1)}$ for any $0\leq j \leq
n-1$, $S$ for storing all the states $|{\rm PA}_n^a(x)\rangle$, and $T$
for storing the $(2^n-1)$-qubit state. All the registers consist of
qubits initialized to $|0\rangle$. The numbers of qubits in $R_j$, $S$,
and $T$ are $2^{n-1}-1$, $2^n-n-1$, and $2^n-1$, respectively. The
circuit is described as follows:
\begin{enumerate}
\item Copy the input state $|x\rangle$ and apply the circuit for
      PA$_n^a$ to each copy for every $a\in \{0,1\}^n \setminus \{0^n\}$
      in parallel.
\begin{enumerate}
\item For each $0\leq j \leq n-1$:

Apply an unbounded fan-out gate to the input qubit in state
      $|x_j\rangle$ and all the qubits in $R_j$, where the input qubit
      is used as the control qubit.

\item For each $0\leq j \leq n-1$:

Apply Hadamard gates to all the qubits in $R_j$.

\item Apply Hadamard gates to all the qubits in $S$.

\item For each $a=a_0\cdots a_{n-1}\in  \{0,1\}^n$ such that $|a| \geq
      2$:

Apply an unbounded fan-out gate to a qubit in $R_{j_0}$, \ldots, a qubit
      in $R_{j_{|a|-1}}$, and a qubit in $S$, where the qubit in $S$ is
      used as the control qubit and $j_0,\cdots,j_{|a|-1}$ is a unique
      sequence of the non-negative integers satisfying $a_{j_0}=\cdots
      =a_{j_{|a|-1}}=1$ and $j_0 < \cdots < j_{|a|-1}$. All the gates
      and the qubits are arranged so that all the gates can be applied
      in parallel.

\item This step is the same as Step~1-(b).

\item This step is the same as Step~1-(c).
\end{enumerate}

\item Apply a Hadamard gate and an unbounded fan-out gate to ancillary
      qubits.
\begin{enumerate}
\item Apply a Hadamard gate to a qubit in $T$.

\item Apply an unbounded fan-out gate to all the qubits in $T$, where
      the qubit to which a Hadamard gate is applied in Step~2-(a) is
      used as the control qubit.
\end{enumerate}

\item Apply controlled-$Z(\pi/2^{n-1})$ gates in parallel to the states
      in Steps~1 and 2.
\begin{enumerate}
\item For each $0\leq j \leq n-1$:

Apply a controlled-$Z(\pi/2^{n-1})$ gate to the input qubit in state
      $|x_j\rangle$ and a qubit in $T$.

\item For each qubit in $S$:

Apply a controlled-$Z(\pi/2^{n-1})$ gate to the qubit in $S$ and a qubit
      in $T$.
\end{enumerate}
All the gates and the qubits are arranged so that all the gates can be
      applied in parallel.

\item Apply an unbounded fan-out gate and a Hadamard gate to the state
      in Step~3.
\begin{enumerate}
\item This step is the same as Step~2-(b).

\item This step is the same as Step~2-(a).
\end{enumerate}
\end{enumerate}
The circuit for $n=3$ is depicted in Fig.~\ref{figure3}. 

The correctness of the circuit is described as follows. Step~1-(a)
transforms the state of $R_j$ into the state $|x_j\rangle^{\otimes
(2^{n-1}-1)}$. Since PA$_n^a(x)$ can be computed by a combination of
Hadamard gates and an unbounded fan-out gate as depicted in
Fig.~\ref{figure2}, Step~1-(f) stores the state $|{\rm
PA}_n^a(x)\rangle$ in $S$ for any $a\in\ \{0,1\}^n$ such that $|a| \geq
2$. Step~2-(a) prepares the state $(H|0\rangle)|0\rangle^{\otimes
(2^n-2)}$ and thus Step~2-(b) transforms the state of $T$ into the
$(2^n-1)$-qubit state
$$\frac{|0\rangle^{\otimes (2^n-1)}+ |1\rangle^{\otimes
(2^n-1)}}{\sqrt{2}}.$$
Step~3 transforms the $(2^n-1)$-qubit state into
$$\frac{|0\rangle^{\otimes (2^n-1)}+ e^{i\pi \frac{1}{2^{n-1}}\sum_{a\in
\{0,1\}^n \setminus \{0^n\}}{\rm PA}_n^a(x)}|1\rangle^{\otimes
(2^n-1)}}{\sqrt{2}},$$
which is equal to
$$\frac{|0\rangle^{\otimes (2^n-1)}+ (-1)^{{\rm
OR}_n(x)}|1\rangle^{\otimes (2^n-1)}}{\sqrt{2}}$$
by Lemma~1. Since Step~4-(a) yields the state
$$\frac{|0\rangle + (-1)^{{\rm OR}_n(x)}|1\rangle}{\sqrt{2}},$$
Step~4-(b) outputs the desired state $|{\rm OR}_n(x)\rangle$.

By the construction, the depth of the whole circuit does not depend on
$n$. Since Step~1-(a) is the dominant part and uses $n$ unbounded
fan-out gates on $2^{n-1}$ qubits, the size of the whole circuit is
$O(n2^n)$. Thus, the depth and size of the whole circuit are $O(1)$ and
$O(n2^n)$, respectively.

\subsection{Proof of Lemma 4}

\begin{figure}[t]
 \vspace{-.3cm}
\centering
\epsfig{file=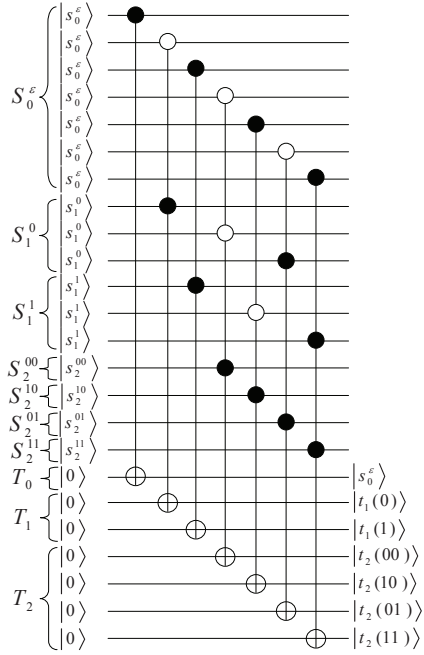,scale=.4}
\vspace{-1.2cm}
\caption{The circuit for Steps~3-(c) and 3-(d) for $m=3$.}
\label{figure4}
\end{figure}

Let $|x\rangle=|x_0\rangle\cdots |x_{n-1}\rangle$ be an input state. As
described in Section~4.1, we prepare the $(2^m-1)$-qubit state
$\bigotimes_{k=0}^{m-1}|\varphi_k\rangle^{\otimes 2^k}$, $2^m-1$ copies
of the state $|s_0^\varepsilon\rangle$ and $2^{m-k}-1$ copies of the
state $|s_k^y\rangle$, and the state $|t_k(y)\rangle$ for any $1 \leq k
\leq m-1$ and $y\in\{0,1\}^k$. Thus, we prepare the registers $R$ for
storing the $(2^m-1)$-qubit state, $S_0^\varepsilon$ for storing the
copies of the state $|s_0^\varepsilon\rangle$, $S_k^y$ for storing the
copies of the state $|s_k^y\rangle$ for any $1 \leq k \leq m-1$ and
$y\in\{0,1\}^k$, $T_0$ for storing the state $|s_0^\varepsilon\rangle$,
and $T_k$ for storing all the states $|t_k(y)\rangle$ for any $1 \leq k
\leq m-1$. All the registers consist of qubits initialized to
$|0\rangle$. The numbers of qubits in $R$, $S_0^\varepsilon$, $S_k^y$,
and $T_k$ are $2^m-1$, $2^m-1$, $2^{m-k}-1$, and $2^k$,
respectively. The circuit is described as follows:
\begin{enumerate}
\item Apply a slightly modified version of H{\o}yer and {\v S}palek's OR
      reduction to the input state $|x\rangle$, where the output is
      stored in $R$.

\item Perform $A(0)$- and
      $A(\pi\sum_{j=0}^{k-1}\frac{y_j}{2^{k-j}})$-measurements for every
      $1 \leq k \leq m-1$ and $y=y_0\cdots y_{k-1}\in \{0,1\}^k$ in
      parallel on the state of $R$.
\begin{enumerate}
\item Perform an $A(0)$-measurement on the state $|\varphi_0\rangle$ of
      $R$ and let $s_0^{\varepsilon}$ be the classical outcome of the
      measurement.

\item For each $1 \leq k \leq m-1$ and $y=y_0\cdots y_{k-1}\in
      \{0,1\}^k$:

Perform an $A(\pi\sum_{j=0}^{k-1}\frac{y_j}{2^{k-j}})$-measurement on
      the state $|\varphi_k\rangle$ of $R$ and let $s_k^y$ be the
      classical outcome of the measurement.
\end{enumerate}

\item Prepare $2^m-1$ copies of the state $|s_0^\varepsilon\rangle$ and
      $2^{m-k}-1$ copies of the state $|s_k^y\rangle$ and apply the
      circuit for AND$_{k+1}$ (constructed by the circuit for OR$_{k+1}$
      in Section~3) to the states for every $1 \leq k \leq m-1$ and $y
      \in \{0,1\}^k$ in parallel.
\begin{enumerate}
\item Apply NOT gates to all the qubits in $S_0^\varepsilon$ if
      $s_0^\varepsilon=1$.

\item For each $1 \leq k \leq m-1$ and $y=y_0\cdots y_{k-1}\in
      \{0,1\}^k$:

Apply NOT gates to all the qubits in $S_k^y$ if $s_k^y=1$.

\item Apply a CNOT gate to a qubit in $S_0^\varepsilon$ and the qubit in
      $T_0$, where the qubit in $S_0^\varepsilon$ is used as the control
      qubit.

\item For each $1 \leq k \leq m-1$ and $y=y_0\cdots y_{k-1}\in
      \{0,1\}^k$:
\begin{itemize}
\item Apply a NOT gate to a qubit (not used in Step~3-(c)) in
      $S_0^\varepsilon$ if $y_0=0$, a NOT gate to a qubit in $S_1^{y_0}$
      if $y_1=0$,\ldots, and a NOT gate to a qubit in
      $S_{k-1}^{y_0\cdots y_{k-2}}$ if $y_{k-1}=0$. All the gates and
      the qubits are arranged so that all the gates can be applied in
      parallel.

\item Apply a gate for AND$_{k+1}$ to the qubit in $S_0^\varepsilon$,
      the qubit in $S_1^{y_0}$, \ldots, the qubit in $S_{k-1}^{y_0\cdots
      y_{k-2}}$, a qubit in $S_k^y$, and a qubit in $T_k$, where the
      output is stored in $T_k$. All the gates and the qubits are
      arranged so that all the gates can be applied in parallel.
\end{itemize}
\end{enumerate}

\item Apply the circuit for PA$_{2^k}$ for every $1 \leq k \leq m-1$ in
      parallel to the state in Step~3.
\begin{enumerate}
\item For each $1 \leq k \leq m-1$:

Apply Hadamard gates to all the qubits in $T_k$.

\item For each $1 \leq k \leq m-1$:

Apply an unbounded fan-out gate to all the qubits in $T_k$.

\item This step is the same as Step~4-(a).
\end{enumerate}
\end{enumerate}
The circuit for Steps~3-(c) and 3-(d) for $m=3$ is depicted in
Fig.~\ref{figure4}. The first half of the whole circuit contains many
one-qubit projective measurements and unitary operations depending on
the classical outcomes of the measurements. We can replace them with
unitary operations including controlled operations and with measurements
in the computational basis only at the end of the circuit by using the
well-known method of coherently implementing measurements
\cite{Nielsen}.

The correctness of the circuit is described as follows. Step~1
transforms the state of $R$ into the state
$\bigotimes_{k=0}^{m-1}|\varphi_k\rangle^{\otimes 2^k}$. Step~2 yields
the values $s_0^\varepsilon,s_k^y\in \{0,1\}$. By the definition of the
measurements, it holds that $s_0=s_0^{\varepsilon}$ and
$s_k=s_k^{s_0\cdots s_{k-1}}$ for any $1 \leq k \leq m-1$. Steps~3-(a)
and 3-(b) transform the state of $S_0^\varepsilon$ into
$|s_0^\varepsilon\rangle^{\otimes(2^m-1)}$ and the state of $S_k^y$ into
$|s_k^y\rangle^{\otimes (2^{m-k}-1)}$ for any $1 \leq k \leq m-1$. Steps
3-(c) and 3-(d) transform the state of $T_0$ into
$|s_0^\varepsilon\rangle$ and the state of $T_k$ into $\bigotimes_{y\in
\{0,1\}^k}|t_k(y)\rangle$. Step~4 
transforms the state of a qubit in $T_k$ into
$|\bigoplus_{y\in\{0,1\}^k}t_k(y)\rangle$ for any $1 \leq k \leq
m-1$. For any $1 \leq k \leq m-1$ and $y \in \{0,1\}^k$,
$$t_k(y) = s_k^y\bigwedge_{j=0}^{k-1}(1\oplus y_j\oplus s_j^{y_0\cdots
      y_{j-1}})
 = s_k^y\bigwedge_{j=0}^{k-1}(1\oplus y_j\oplus s_j^{s_0\cdots
      s_{j-1}})
 = s_k^y\bigwedge_{j=0}^{k-1}(1\oplus y_j\oplus s_j)$$
and thus
$$t_k(y)= \left\{
\begin{array}{cc}
s_k, & \mbox{if $y=s_0\cdots s_{k-1}$,} \\
0,   & \mbox{otherwise.}
\end{array}
\right.$$
Therefore, $\bigoplus_{y\in\{0,1\}^k}t_k(y)=s_k$ for any $1 \leq k \leq
m-1$. Thus, Step~4 outputs the desired state $|s_k\rangle$ for any $1
\leq k \leq m-1$.

By the construction, the depth of the whole circuit does not depend on
$n$. Since Step~1 is the dominant part and the state
$\bigotimes_{k=0}^{m-1}|\varphi_k\rangle^{\otimes 2^k}$ in Step~1 can be
prepared with a circuit of size $O(n\sum_{k=0}^{m-1}2^k)=O(n^2)$ as in
H{\o}yer and {\v S}palek's OR reduction, the size of the whole circuit
is $O(n^2)$. Therefore, the depth and size of the whole circuit are
$O(1)$ and $O(n^2)$, respectively.

\subsection{Proof of Lemma 5}

Let $t$ be an integer satisfying $1 \leq t \leq \lceil n/2 \rceil$ and
$|x\rangle = |x_0\rangle\cdots|x_{n-1}\rangle$ be an input state. Let
$l$ be an integer satisfying $0 \leq l < \lceil \log (t+1) \rceil$. This
means that $l$ is less than the length of the binary representation of
$t$. Let $t_0 \cdots t_{l-1}$ be the $l$ low-order bits of the binary
representation of $t$, where $t_0$ is the lowest-order bit. Note that
the value $t-\sum_{j=0}^{l-1}t_j2^j$ is positive and is a multiple of
$2^l$. The first circuit is described as follows:
\begin{enumerate}
\item  Apply the circuit in Lemma~4 to the input state $|x\rangle$,
       where we regard $m$ in the proof of Lemma~4 as $l$. Let
       $|s_0\rangle \cdots |s_{l-1}\rangle$ be the output. In other
       words, $s_0\cdots s_{l-1}$ are the $l$ low-order bits of the
       binary representation of $|x|$, where $s_0$ is the lowest-order
       bit.

\item  Apply the first circuit in Lemma~3 to the input state
       $|x\rangle$, where we consider only $0 \leq k \leq t-1$ such that
       the $l$ low-order bits of the binary representation of $k$ are
       equal to $s_0\cdots s_{l-1}$. More concretely,
$$k=M 2^l + \sum_{j=0}^{l-1}s_j 2^j$$
for any integer $M$ satisfying
$$0 \leq M \leq \frac{t-\sum_{j=0}^{l-1}t_j2^j}{2^l}$$
if $\sum_{j=0}^{l-1}t_j2^j >  \sum_{j=0}^{l-1}s_j2^j$ and
$$0 \leq M \leq \frac{t-\sum_{j=0}^{l-1}t_j2^j}{2^l}-1$$
otherwise.
\end{enumerate}
We note that, before Step~2, we prepare all the binary representations
of $k$ satisfying the above conditions by applying unbounded fan-out
gates and NOT gates to ancillary qubits (initialized to $|0\rangle$).

As in the proof of Lemma~3, the circuit outputs the desired state $|{\rm
TH}_n^t(x)\rangle$ and the depth of the whole circuit does not depend on
$n$. The sizes of the circuits in Steps~1 and 2 are $O(2^ln)$ and
$O(2^{-l}tn\log n)$, respectively, since $M \leq t/2^l$. Thus, the depth
and size of the whole circuit are $O(1)$ and $O(2^ln+2^{-l}tn\log n)$,
respectively. To construct the second circuit, we use the second circuit
in Lemma~3, where we consider only $t \leq k \leq n$ such that the $l$
low-order bits of the binary representation of $k$ are $s_0\cdots
s_{l-1}$. The number of $k$'s we need to consider is bounded above by
$(n-t+1)/2^l+2$ and thus the depth and size of the resulting circuit are
$O(1)$ and $O(2^ln+2^{-l}(n-t+1)n\log n+n\log n)$, respectively.

\subsection{Proof of Theorem~2}

For any $1 \leq t \leq \log n$, it holds that $0 \leq \lceil \log (t+1)
\rceil -1 < \lceil \log (t+1) \rceil$ and thus we set $l=\lceil \log
(t+1) \rceil -1$ in the first circuit in Lemma~5. This yields an
$O(n\log n)$-size circuit. For any $\log n \leq t \leq \lceil n/2
\rceil$, it holds that $0 \leq \lceil\log\sqrt{t\log n}\rceil -1 <
\lceil \log (t+1) \rceil$ and thus we set $l=\lceil \log\sqrt{t\log
n}\rceil -1$ in the first circuit in Lemma~5. This yields an
$O(n\sqrt{t\log n})$-size circuit. For any $\lceil n/2 \rceil \leq t
\leq n-\log n$, it holds that $0 \leq \lceil\log\sqrt{(n-t+1)\log
n}\rceil -1 < \lceil \log (t+1) \rceil$ and thus we set $l=\lceil
\log\sqrt{(n-t+1)\log n}\rceil -1$ in the second circuit in
Lemma~5. This yields an $O(n\sqrt{(n-t)\log n})$-size circuit. For any
$n-\log n \leq t \leq n$, it holds that $0 \leq \lceil\log(n-t+2)\rceil
-1 < \lceil \log (t+1) \rceil$ and thus we set $l=\lceil
\log(n-t+2)\rceil -1$ in the second circuit in Lemma~5. This yields an
$O(n\log n)$-size circuit.

\subsection{Proof of Theorem~3}

As described in Section~5, it suffices to show that, if F$_p$ is in $\qnc$, 
there exists a poly$(n)$-time exact classical algorithm for the DLP over 
$G$ using the $\qnc$ oracle. We consider a slightly modified version of 
van Dam's exact algorithm for the DLP. The main difference is that 
the slightly modified version does not include intermediate 
measurements. This allows us to consider an exact algorithm with a 
simple structure: a poly$(n)$-time classical pre-processing, a query 
to the $\qnc$ oracle, and a poly$(n)$-time classical post-processing.

Let $x \equiv g^l \in G$ $(0 \leq l \leq p-1)$ be an input. In the 
classical pre-processing step, we compute the values 
$g^{2^j} \ {\rm mod} \ q$, $x^{2^j} \ {\rm mod} \ q$, 
and $(x^{-1})^{2^j} \ {\rm mod} \ q$ $(0 \leq j \leq m-1)$ by a 
poly$(n)$-time exact classical algorithm. By a query to the $\qnc$ oracle, 
we solve the problem of finding $s$ and $sl \ {\rm mod} \ p$ for some 
$1 \leq s \leq p-1$ using the pre-computed values. In the 
classical post-processing step, using the values $s$ and 
$sl \ {\rm mod} \ p$ obtained from the $\qnc$ oracle, we compute 
$sl\cdot s^{-1} \ {\rm mod} \ p = l$, which is the desired 
output, by a poly$(n)$-time exact classical algorithm. This can always 
be done since $\gcd(s,p)=1$ for any $1 \leq s \leq p-1$. Thus, the only 
problem is to show that the $\qnc$ oracle can solve the problem, 
in other words, to show that the problem can be solved exactly by a 
$\qnc$ circuit (if F$_p$ is in $\qnc$).

The quantum algorithm for solving the problem consists of two parts 
$Q_1$ and $Q_2$. We note that we can use the pre-computed values 
descried above in the quantum algorithm. The first part $Q_1$ 
transforms the state $|0\rangle^{\otimes (m+n+1)}$ into the state
$$\frac{1}{\sqrt{p-1}} \sum_{s=1}^{p-1}|s\rangle |\chi^s\rangle 
|1\rangle,$$
which is independent of the input $x$. To define $Q_1$, we define 
the following algorithm as $\cal A$, where the input state 
is $|0\rangle^{\otimes (m+n+1)}$:
\begin{enumerate}
\item Apply F$_p$ to the first $m$ qubits of the input state.

\item Apply the modular exponentiation operation $|r\rangle|0\rangle \to
      |r \rangle |g^r \ {\rm mod} \ q \rangle$ to the state in Step 1.
     
\item Apply F$_p$ to the first $m$ qubits of the state in Step 2.

\item Apply the one-qubit unitary operation defined by
$$\frac{1}{\sqrt{2(p-1)}}
\bigg(
\begin{array}{cc}
\sqrt{p-2} & -\sqrt{p} \\
\sqrt{p} & \sqrt{p-2}
\end{array}
\bigg)$$
to the last one qubit of the state in Step~3.
\end{enumerate}

A direct calculation shows that $\cal A$ transforms the input state into
\begin{eqnarray*}
&& \frac{1}{\sqrt{p}} \sum_{s=0}^{p-1}|s\rangle |\chi^s\rangle 
\left(\sqrt{\frac{p-2}{2(p-1)}}|0\rangle + 
\sqrt{\frac{p}{2(p-1)}}|1\rangle)\right)\\
&=& \frac{1}{\sqrt{2(p-1)}} \sum_{s=1}^{p-1}|s\rangle |\chi^s\rangle
 |1\rangle
+ \frac{1}{\sqrt{2(p-1)}} |0\rangle |\chi^0\rangle |1\rangle
+ \sqrt{\frac{p-2}{2p(p-1)}} \sum_{s=0}^{p-1}|s\rangle |\chi^s\rangle
|0\rangle.
\end{eqnarray*}
We define
$$|A\rangle = \frac{1}{\sqrt{2(p-1)}} \sum_{s=1}^{p-1}|s\rangle
|\chi^s\rangle |1\rangle.$$
It holds that $\langle A|A\rangle=1/2$. Let $S_{\{0\}}$ be the
quantum operation that changes the phase of a state by $i$ if and only
if the state is $|0\rangle^{\otimes (m+n+1)}$. Similarly, let $S_A$ be
the quantum operation that changes the phase of a state by $i$ if and
only if the state of the first $m$ qubits is not $|0\rangle^{\otimes m}$ 
and the state of the last one qubit is $|1\rangle$. We define the 
Grover iteration $G={\cal A}S_{\{0\}}{\cal A}^{-1}S_A$ and the first 
part $Q_1=G{\cal A}$. The correctness of $Q_1$ follows from the 
direct calculation as in the amplitude amplification procedure 
in~\cite{Brassard0,Brassard}.

The argument in Section 5 implies that $\cal A$ is 
in $\qnc$. Moreover, by Theorem~1, $S_{\{0\}}$ and $S_A$ are in $\qnc$. 
Thus, $Q_1$ is in $\qnc$. We note that the last qubit in state
$|1\rangle$ is not important for the second part $Q_2$ describe below 
(and thus can be ignored below) and that the pre-computed values 
used in Step~2 on ancillary qubits have no effect on the 
amplitude amplification procedure.

Recall that the quantum operation $D_x$ is defined as
$$|y\rangle|\alpha\rangle \mapsto |y\cdot x^{-\alpha}
	  \ {\rm mod} \ q\rangle|\alpha\rangle,$$
where $0 \leq y \leq q-1$ and $0 \leq \alpha \leq p-1$. Before
considering the second part $Q_2$, we show that the relationship
$$D_x|\chi^s\rangle|\alpha\rangle = \omega_p^{sl\alpha} |\chi^s\rangle
|\alpha\rangle$$
holds for any $0 \leq s \leq p-1$ and $0 \leq \alpha \leq p-1$ by the
following direct calculation:
\begin{eqnarray*}
D_x|\chi^s\rangle|\alpha\rangle &=&
 \frac{1}{\sqrt{p}}\sum_{r=0}^{p-1}\omega _p^{sr}D_x|g^r \  {\rm mod} \
 q\rangle |\alpha\rangle
=
 \frac{1}{\sqrt{p}}\sum_{r=0}^{p-1}\omega _p^{sr}|g^r\cdot
 x^{-\alpha} \  {\rm mod} \ q\rangle |\alpha\rangle\\
&=&
\omega_p^{sl\alpha} \frac{1}{\sqrt{p}}\sum_{r=0}^{p-1}\omega _p^{s(r-l
\alpha)}|g^{r-l \alpha} \  {\rm mod} \ q\rangle |\alpha\rangle
=
\omega_p^{sl\alpha} |\chi^s\rangle |\alpha\rangle.
\end{eqnarray*}

We consider the second part $Q_2$ that transforms the input state
$$\frac{1}{\sqrt{p-1}} \sum_{s=1}^{p-1}|s\rangle |\chi^s\rangle
|0\rangle^{\otimes m},$$
which is obtained by $Q_1$ with $m$ qubits initialized to $|0\rangle$, 
into the state 
$$\frac{1}{\sqrt{p-1}} \sum_{s=1}^{p-1}|s\rangle |\chi^s\rangle
	  |sl \ {\rm mod} \ p\rangle.$$
We define the following algorithm as $Q_2$:
\begin{enumerate}
\item[5.] Apply F$_p$ to the last $m$ qubits of the input state.

\item[6.] Apply $D_x$ to the last $n+m$ qubits of the state in Step~5.

\item[7.] Apply F$_p^{-1}$ to the last $m$ qubits of the state in Step~6.
\end{enumerate}

The correctness of $Q_2$ is described as follows. Step~5 transforms 
the input state into the state
$$\frac{1}{\sqrt{p-1}} \sum_{s=1}^{p-1}|s\rangle
	  |\chi^s\rangle\left(\frac{1}{\sqrt{p}} \sum_{\alpha=0}^{p-1}
	  |\alpha\rangle\right).$$
By the relationship shown above, Step~6 transforms the state in Step~5 
into the state 
$$\frac{1}{\sqrt{p-1}} \sum_{s=1}^{p-1}|s\rangle
	  |\chi^s\rangle \left(\frac{1}{\sqrt{p}} \sum_{\alpha=0}^{p-1}
	  \omega_p^{sl\alpha}|\alpha\rangle\right).$$
Step~7 transforms the state in Step~6 into the desired state
$$\frac{1}{\sqrt{p-1}} \sum_{s=1}^{p-1}|s\rangle |\chi^s\rangle
	  |sl \ {\rm mod} \ p\rangle.$$
We perform one-qubit projective measurements in the basis 
$|0\rangle,|1\rangle$ on the first $m$ qubits and the last $m$ qubits of 
the state in Step~7. This yields the classical outcomes $s$ and 
$sl \ {\rm mod} \ p$ for some $1 \leq s \leq p-1$. 

Steps 5 and 7 are in $\qnc$ by our assumption. In Step~6, as in Step~2 of 
$\cal A$, $D_x$ is implemented by using the iterated 
multiplication operation with the pre-computed values 
($x^{2^j} \ {\rm mod} \ q$ and $(x^{-1})^{2^j} \ {\rm mod} \ q$) and 
the modular multiplication operation as follows:
\begin{eqnarray*}
|y\rangle|\alpha\rangle |0\rangle^{\otimes 2n} &\mapsto&
 |y\rangle|\alpha\rangle |x^{-\alpha} \ {\rm mod} \
 q\rangle|0\rangle^{\otimes
 n}\\
&\mapsto&
 |y\rangle|\alpha\rangle |x^{-\alpha} \ {\rm mod} \
 q\rangle|y\cdot x^{-\alpha} \ {\rm mod} \ q\rangle\\
&\mapsto&
 |y\rangle|\alpha\rangle |0\rangle^{\otimes n}|y\cdot x^{-\alpha} \ {\rm
 mod} \ q\rangle\\
&\mapsto&
 |y\rangle|\alpha\rangle |x^\alpha \ {\rm mod} \ q\rangle|y\cdot
 x^{-\alpha} \ {\rm mod} \ q\rangle\\
&\mapsto&
 |0\rangle^{\otimes n}|\alpha\rangle |x^\alpha \ {\rm mod} \
 q\rangle|y\cdot x^{-\alpha} \ {\rm mod} \ q\rangle\\
&\mapsto&
 |0\rangle^{\otimes n}|\alpha\rangle |0\rangle^{\otimes n}|y\cdot
 x^{-\alpha} \ {\rm mod} \ q\rangle.
\end{eqnarray*}
Since $\qnc = \qtc$ as shown in Section~3 and $\qtc$ includes the 
iterated multiplication operation and the modular multiplication 
operation \cite{Siu}, Step 6 is in $\qnc$. Therefore, $Q_2$ is in $\qnc$.


\begin{thebibliography}{99}

\bibitem{Aaronson}
Aaronson, S.: BQP and the polynomial hierarchy, ACM Symposium on Theory
	of Computing, 141--150 (2010).

\bibitem{Bera}
Bera, D., Green, F., Homer, S.: Small depth quantum circuits, ACM SIGACT
	NEWS 38 (2), 35--50 (2007).

\bibitem{Bera2}
Bera, D.: A lower bound method for quantum circuits, Information
	Processing Letters 111 (15), 723--726 (2011).

\bibitem{Brassard0}
Brassard, G., H\o yer, P.: An exact quantum polynomial-time algorithm for
	Simon's problem, Israeli Symposium on Theory of Computing and
	Systems, 12--23 (1997).

\bibitem{Brassard}
Brassard, G., H\o yer, P., Mosca, M., Tapp, A.: Quantum amplitude
	amplification and estimation, Quantum Computation and Quantum
	Information: A Millennium Volume, AMS Contemporary Mathematics
	Series 305, 53--74 (2002).

\bibitem{Browne}
Browne, D.E., Kashefi, E., Perdrix, S.: Computational depth complexity
	of measurement-based quantum computation, Conference on Theory
	of Quantum Computation, Communication, and Cryptography 2010,
	LNCS 6519, 35--46 (2011).

\bibitem{Chandra}
Chandra, A.K., Fortune, S., Lipton, R.: Unbounded fan-in circuits and
	associative functions, ACM Symposium on Theory of Computing,
	52--60 (1983).

\bibitem{Cleve}
Cleve, R., Watrous, J.: Fast parallel circuits for the quantum Fourier
	transform, IEEE Symposium on Foundations of Computer Science,
	526--536 (2000).

\bibitem{Fang}
Fang, M., Fenner, S., Green, F., Homer, S., Zhang, Y.: Quantum lower
	bounds for fanout, Quantum Information and Computation 6 (1),
	46--57 (2006).

\bibitem{Fenner}
Fenner, S., Green, F., Homer, S., Zhang, Y.: Bounds on the power of
	constant-depth quantum circuits, Fundamentals of Computation
	Theory, LNCS 3623, 44--55 (2005).

\bibitem{Furst}
Furst, M., Saxe, J.B., Sipser, M.: Parity, circuits, and the polynomial
	hierarchy, Mathematical Systems Theory 17, 13--27 (1984).

\bibitem{Green}
Green, F., Homer, S., Moore, C., Pollett, C.: Counting, fanout, and the
	complexity of quantum ACC, Quantum Information and Computation
	2 (1), 35--65 (2002).

\bibitem{Hales}
Hales, L., Hallgren, S.: An improved quantum Fourier transform algorithm
	and applications, IEEE Symposium on Foundations of Computer
	Science, 515--525 (2000).

\bibitem{Hoban}
Hoban, M.J., Campbell, E.T., Loukopoulos, K., Browne, D.E.: Non-adaptive
	measurement-based quantum computation and multi-party Bell
	inequalities, New Journal of Physics 13, 023014 (2011).

\bibitem{Hoyer}
H{\o}yer, P., {\v S}palek, R.: Quantum fan-out is powerful, Theory of
	Computing 1 (5), 81--103 (2005).

\bibitem{Moore}
Moore, C., Nilsson, M.: Parallel quantum computation and quantum codes,
	SIAM Journal on Computing 31 (3), 799--815 (2001).

\bibitem{Mosca}
Mosca, M., Zalka, Ch.: Exact quantum Fourier transforms and discrete
	logarithm algorithms, International Journal of Quantum
	Information 2 (1), 91--100 (2004).

\bibitem{Nielsen}
Nielsen, M.A., Chuang, I.L.: Quantum Computation and Quantum
	Information, Cambridge University Press (2000).

\bibitem{Odonnell}
O'Donnell, R.: Some topics in analysis of Boolean functions, ACM
	Symposium on Theory of Computing, 569--578 (2008).

\bibitem{Pohlig}
Pohlig, S.C., Hellman, M.E.: An improved algorithm for computing
	logarithms over GF$(p)$ and its cryptographic significance, IEEE
	Transactions on Information Theory 24 (1), 106--110 (1978).

\bibitem{Shor}
Shor, P.W.: Polynomial-time algorithms for prime factorization and
	discrete logarithms on a quantum computer, SIAM Journal on
	Computing 26 (5), 1484--1509 (1997).

\bibitem{Siu}
Siu, K.-Y., Bruck, J., Kailath, T., Hofmeister, T.: Depth efficient
	neural networks for division and related problems, IEEE
	Transactions on Information Theory 39 (3), 946--956 (1993).

\bibitem{van}
van Dam, W.: Quantum computing discrete logarithms with the help of a
	preprocessed state, arXiv:quant-ph/0311134.

\bibitem{Vollmer}
Vollmer, H.: Introduction to Circuit Complexity, Springer (1999).

\end{thebibliography}
\end{document}